

\message
{JNL.TEX version 0.92 as of 6/9/87.  Report bugs and problems to Doug Eardley.}

\catcode`@=11
\expandafter\ifx\csname inp@t\endcsname\relax\let\inp@t=\input
\def\input#1 {\expandafter\ifx\csname #1IsLoaded\endcsname\relax
\inp@t#1%
\expandafter\def\csname #1IsLoaded\endcsname{(#1 was previously loaded)}
\else\message{\csname #1IsLoaded\endcsname}\fi}\fi
\catcode`@=12







\def\beginlinemode{\endmode
  \begingroup\parskip=0pt \obeylines\def\\{\par}\def\endmode{\par\endgroup}}
\def\beginparmode{\endmode
  \begingroup \def\endmode{\par\endgroup}}
\let\endmode=\par
{\obeylines\gdef\
{}}
\def\singlespace{\baselineskip=\normalbaselineskip}

\def\oneandahalfspace{\baselineskip=\normalbaselineskip
  \multiply\baselineskip by 3 \divide\baselineskip by 2}
\def\doublespace{\baselineskip=\normalbaselineskip \multiply\baselineskip by 2}






\def\title			
  {\null\vskip 3pt plus 0.2fill
   \beginlinemode \doublespace \raggedcenter \bf}

\def\author			
  {\vskip 3pt plus 0.2fill \beginlinemode
   \singlespace \raggedcenter\sc}

\def\affil			
  {\vskip 3pt plus 0.1fill \beginlinemode
   \oneandahalfspace \raggedcenter \sl}

\def\abstract			
  {\vskip 3pt plus 0.3fill \beginparmode
   \oneandahalfspace ABSTRACT: }

\def\endtitlepage		
  {\endpage			
   \body}

\def\body			
  {\beginparmode}		

\def\beginitems{
\par\medskip\bgroup\def\i##1 {\item{##1}}\def\ii##1 {\itemitem{##1}}
\leftskip=36pt\parskip=0pt}
\def\enditems{\par\egroup}

\def\beneathrel#1\under#2{\mathrel{\mathop{#2}\limits_{#1}}}

\def\refto#1{~[{#1}]}

\def\references			
  {
   \beginparmode
   \frenchspacing \parindent=0pt \leftskip=1truecm
   \parskip=8pt plus 3pt \everypar{\hangindent=\parindent}}

\gdef\refis#1{\item{#1.\ }}			

\gdef\journal#1, #2, #3, 1#4#5#6{		
    {\sl #1~}{\bf #2}, #3 (1#4#5#6)}		

\def\endreferences{\body}

\catcode`@=11
\newcount\r@fcount \r@fcount=0
\newcount\r@fcurr
\immediate\newwrite\reffile
\newif\ifr@ffile\r@ffilefalse
\def\w@rnwrite#1{\ifr@ffile\immediate\write\reffile{#1}\fi\message{#1}}

\def\writer@f#1>>{}
\def\referencefile{
  \r@ffiletrue\immediate\openout\reffile=\jobname.ref%
  \def\writer@f##1>>{\ifr@ffile\immediate\write\reffile%
    {\noexpand\refis{##1} = \csname r@fnum##1\endcsname = %
     \expandafter\expandafter\expandafter\strip@t\expandafter%
     \meaning\csname r@ftext\csname r@fnum##1\endcsname\endcsname}\fi}%
  \def\strip@t##1>>{}}

\def\citeall#1{\xdef#1##1{#1{\noexpand\cite{##1}}}}
\def\cite#1{\each@rg\citer@nge{#1}}	

\def\each@rg#1#2{{\let\thecsname=#1\expandafter\first@rg#2,\end,}}
\def\first@rg#1,{\thecsname{#1}\apply@rg}	
\def\apply@rg#1,{\ifx\end#1\let\next=\relax
\else,\thecsname{#1}\let\next=\apply@rg\fi\next}

\def\citer@nge#1{\citedor@nge#1-\end-}	
\def\citer@ngeat#1\end-{#1}
\def\citedor@nge#1-#2-{\ifx\end#2\r@featspace#1 
  \else\citel@@p{#1}{#2}\citer@ngeat\fi}	
\def\citel@@p#1#2{\ifnum#1>#2{\errmessage{Reference range #1-#2\space is bad.}%
    \errhelp{If you cite a series of references by the notation M-N, then M and
    N must be integers, and N must be greater than or equal to M.}}\else%
 {\count0=#1\count1=#2\advance\count1
by1\relax\expandafter\r@fcite\the\count0,%
  \loop\advance\count0 by1\relax
    \ifnum\count0<\count1,\expandafter\r@fcite\the\count0,%
  \repeat}\fi}

\def\r@featspace#1#2 {\r@fcite#1#2,}	
\def\r@fcite#1,{\ifuncit@d{#1}
    \newr@f{#1}%
    \expandafter\gdef\csname r@ftext\number\r@fcount\endcsname%
                     {\message{Reference #1 to be supplied.}%
                      \writer@f#1>>#1 to be supplied.\par}%
 \fi%
 \csname r@fnum#1\endcsname}
\def\ifuncit@d#1{\expandafter\ifx\csname r@fnum#1\endcsname\relax}%
\def\newr@f#1{\global\advance\r@fcount by1%
    \expandafter\xdef\csname r@fnum#1\endcsname{\number\r@fcount}}

\let\r@fis=\refis			
\def\refis#1#2#3\par{\ifuncit@d{#1}
   \newr@f{#1}%
   \w@rnwrite{Reference #1=\number\r@fcount\space is not cited up to now.}\fi%
  \expandafter\gdef\csname r@ftext\csname r@fnum#1\endcsname\endcsname%
  {\writer@f#1>>#2#3\par}}

\def\ignoreuncited{
   \def\refis##1##2##3\par{\ifuncit@d{##1}%
     \else\expandafter\gdef\csname r@ftext\csname
r@fnum##1\endcsname\endcsname%
     {\writer@f##1>>##2##3\par}\fi}}

\def\r@ferr{\endreferences\errmessage{I was expecting to see
\noexpand\endreferences before now;  I have inserted it here.}}
\let\r@ferences=\references
\def\references{\r@ferences\def\endmode{\r@ferr\par\endgroup}}

\let\endr@ferences=\endreferences
\def\endreferences{\r@fcurr=0
  {\loop\ifnum\r@fcurr<\r@fcount
    \advance\r@fcurr by 1\relax\expandafter\r@fis\expandafter{\number\r@fcurr}%
    \csname r@ftext\number\r@fcurr\endcsname%
  \repeat}\gdef\r@ferr{}\endr@ferences}


\let\r@fend=\endpaper\gdef\endpaper{\ifr@ffile
\immediate\write16{Cross References written on []\jobname.REF.}\fi\r@fend}

\catcode`@=12

\citeall\refto		


\magnification 1200
\vsize=7.8in
\hsize=5.7in
\voffset=0.1in
\hoffset=-0.2in
\newcount\eqnumber
\baselineskip 18pt plus 0pt minus 0pt


\font\rmmthree=cmbx10 scaled 1500
\font\rmmtwo=cmbx10 scaled 1200
\font\rmmoneB=cmbx10 scaled 1100

\font\rmmoneI=cmti10 scaled 1000

\font\ninerm=cmr10 scaled 900

\font\eightit=cmti10 scaled 800
\font\eightrm=cmr10 scaled 800
\font\eightbf=cmbx10 scaled 800


\def\title#1{\centerline{\noindent{\rmmthree #1}}\nobreak\medskip\eqnumber=1}

\def\sectbegin#1#2{\bigskip\bigbreak\leftline{\rmmtwo
#1~#2}\nobreak\medskip\nobreak}

\def\nosectbegin#1{\bigskip\bigbreak\leftline{\rmmtwo #1}\nobreak\medskip}


\def\lapp{\hbox{$ {     \lower.40ex\hbox{$<$}
                   \atop \raise.20ex\hbox{$\sim$}
                   }     $}  }
\def\gapp{\hbox{$ {     \lower.40ex\hbox{$>$}
                   \atop \raise.20ex\hbox{$\sim$}
                   }     $}  }

\def\marbul{\strut\vadjust{\kern-2pt$\bullet$}}

\def\specialwarn{\vtop to
\strutdepth{\baselineskip\strutdepth\vss\llap{
\lower.1ex\hbox{$\bigtriangleup$}\kern-0.884em$\triangle$\kern-0.5667em{\eightrm
!}\hskip 13.5pt}\null}}
\def\strutdepth{\dp\strutbox}


\def\new{{\the\eqnumber}\global\advance\eqnumber by 1}
\def\delaynew{{\the\eqnumber}}
\def\nownew{\global\advance\eqnumber by 1}
\def\last{\advance\eqnumber by -1 {\the\eqnumber}
    \global\advance\eqnumber by 1}
\def\eqnam#1{
\xdef#1{\the\eqnumber}}


\def\caption#1#2{
\baselineskip 10pt\noindent\narrower\rm\hbox{\eightbf
#1}:\quad\eightrm
#2 \smallskip}

\def\picture #1 by #2 (#3){
  \vbox to #2{
    \hrule width #1 height 0pt depth 0pt
    \vfill
    \special{picture #3} 
    }
  }

\def\scaledpicture #1 by #2 (#3 scaled #4){{
  \dimen0=#1 \dimen1=#2
  \divide\dimen0 by 1000 \multiply\dimen0 by #4
  \divide\dimen1 by 1000 \multiply\dimen1 by #4
  \picture \dimen0 by \dimen1 (#3 scaled #4)}
  }


\def\dalemb#1#2{{\vbox{\hrule height .#2pt
\hbox{\vrule width.#2pt height#1pt \kern#1pt\vrule width.#2pt}
\hrule height.#2pt}}}

\def\tdot{\kern -8.5pt {}^{{}^{\hbox{...}}}}
\def\dotprime{\kern -8.0pt{}^{{}^{\hbox{.}~\prime}}}
\def\ov{\overline}
\def\dL{\delta_{_{\rm L}}}
\def\dE{\delta_{_{\rm E}}}

\def\nabl{\nabla} 
\def\R{{\cal R}}
\def\I{{\cal I}}
\def\L{{\cal L}} \def\Lg{{\cal L}_{_{\rm\bf G}}}
\def\Tg{T_{\!\!_{\rm\bf G}}\!}
\def\C{{\cal C}}
\def\sp{\ \,}


\title{GRAVITATIONAL PERTURBATIONS OF}
\title{RELATIVISTIC MEMBRANES AND STRINGS}
\medskip
\centerline{\rmmoneB R.$\,$A. Battye}
\vskip 6pt
\baselineskip 12pt
\centerline{\rmmoneI Department of Applied Mathematics and Theoretical Physics}

\centerline{\rmmoneI University of Cambridge}

\centerline{{\rmmoneI Silver Street, Cambridge~CB3 9EW, U.K.}
\footnote{*}{\noindent Email: rab17$\,$@$\,$damtp.cam.ac.uk}}
\medskip
\centerline{and}
\medskip
\centerline{\rmmoneB B. Carter}
\vskip 6pt
\baselineskip 12pt
\centerline{\rmmoneI Department d'Astrophysique Relativiste et de Cosmologie}

\centerline{\rmmoneI Centre National de la Recherche Scientifique, Observatoire
de Paris}

\centerline{{\rmmoneI 92195 Meudon Cedex, France.}
\footnote{\dag}{{\noindent Email: carter$\,$@$\,$mesiob.obspm.fr}\smallskip
\indent Paper submitted to {\eightit Physics Letters} {\eightbf
B}.\smallskip}}
\bigskip \centerline{\rmmoneB Abstract}

\medskip {\narrower{\baselineskip 9pt \ninerm \noindent We consider
gravitationally induced perturbations of relativistic  Dirac--Goto--Nambu
membranes and strings (or {\it p}-branes). The dynamics are described by
the first and second fundamental tensors, and related curvature tensors in an
{\it n}-dimensional spacetime. We show how equations of motion can be
derived for the perturbations within a general gauge and then discuss how
various simple gauge choices can be used to simplify the equations of motion
for specific applications. We also show how the same equations of motion can be
derived from an effective action by a variational principle. Finally, we
compare these equations of motion to those using more familiar notation for
brane dynamics, which involves the induced metric on the worldsheet. This work
sets up a general formalism for understanding the effects of backreaction on
brane dynamics and the background curvature.\smallskip}}

\smallskip \baselineskip 18pt\tenrm

\sectbegin{1.}{Introduction}

\noindent Relativistic membranes and strings (or {\it p}-branes) occur as
topological defects and other solitonic structures in a variety of physical
contexts\refto{NATOVS}. Possibly the most exciting of these is the
formation of defects during phase transitions in the early
universe\refto{kibble}. The localised energy of these defects is likely to
extremely large and therefore their gravitational effects  maybe cosmologically
significant. In particular, cosmic strings may have been  the initial seeds for
the formation of galaxies and other large scale structure\refto{structure}.
Therefore, an understanding of the precise dynamics of branes is
of significant interest. Of particular interest is effect of backreaction on
the dynamics of the brane and the related effects on background curvature. In
this letter we set up a mathematical formalism by which such effects can be
studied. The original motivation was to study the effects of gravitational
radiation backreaction\refto{BSa,Bata} on a network of cosmic
strings\refto{ASa}
and subsequent stochastic gravitational radiation background\refto{BB,ASb,AC}.
However, the results presented here are completely general and  apply to a {\it
p}-dimensional brane in an {\it n}-dimensional spacetime. (NB. A 0--brane is a
point particle, a 1--brane is a string, etc.)

The usual approach to brane dynamics involves specifying the coordinates
of the brane $X^{\mu}=X^{\mu}(\sigma^a)$, where $a=0,..,p$ and the $\sigma^a$
are internal coordinates on the worldsheet. The spacetime interval between two
neighbouring points is
\eqnam{\interval}
$$ds^2=g_{\mu\nu}\partial_aX^\mu\partial_bX^{\nu}d\sigma^ad\sigma^b
\,, \eqno(\new)$$
where $g_{\mu\nu}$ is the spacetime metric and
$\partial_a=\partial/\partial\sigma^{a}$. Hence, the induced
{\it p}+1 dimensional worldsheet metric is given by  \eqnam{\worldmet}
$$\gamma_{ab}=g_{\mu\nu}\partial_aX^{\mu}\partial_bX^{\nu}\,.\eqno(\new)$$
The contravariant inverse metric tensor $\gamma^{ab}$ can be defined as
usual by $\gamma^{ab}\gamma_{bc}=\delta^{a}_{~c}$ and
$\Vert\gamma\Vert=\hbox{det}(\gamma_{ab})$ is the determinant of the
induced metric. This bi-tensorial approach, using quantities such as
$\partial_aX^\mu$ that are tensorial with respect to both background and
internal indices, is useful for explicit computation in physical applications.
However,  long calculations using this traditional approach can become
extremely cumbersome  due to the large number of internal indices involved. For
this reason, we  will use the more concise pure background tensorial
formalism for brane dynamics developed in refs.\refto{car1,car2,car3,CMS}. In
this formalism calculations are simplified by the lack of dependence on gauge
and internal coordinate choices. Simple gauge and coordinate choices will allow
one to convert the expressions deduced here into more physically usable
expressions.

In the application of this formalism it is desirable  to organise the
tensors governing the dynamics  in terms of components
that are tangential or perpendicular to the worldsheet. To this end
we define the first fundamental tensor, or tangential projection tensor, as
\eqnam{\firstten}
$$\eta^{\mu\nu}=\gamma^{ab} \partial_aX^{\mu}\partial_bX^{\nu}\,,\eqno(\new)$$
and we use the notation
$\perp^{\!\mu\nu}=g^{\mu\nu}-\eta^{\mu\nu}$ for its orthogonal complement.
 For tensor fields whose
support is confined to the worldsheet only the tangentially projected covariant
differentiation,
\eqnam{\covdiff}
$$\ov\nabla_{\mu}=\eta_{\mu}^{\sp\nu}\nabla_{\nu}\,,\eqno(\new)$$
is well defined. Using this one can define the second fundamental tensor and
the curvature vector as \eqnam{\secten}
$$K_{\mu\nu}^{\sp\sp\rho}=\eta_{\sigma\nu}\ov\nabla_{\mu}\eta^{\sigma\rho}\,,
\quad\quad K^{\rho}=g^{\mu\nu}K_{\mu\nu}^{\sp\sp\rho}\,.\eqno(\new)$$
The second fundamental tensor can be shown to have the following
elementary properties
\eqnam{\tenprop}
$$\perp^{\sigma\mu}K_{\mu\nu}^{\sp\sp
\rho}=0\,,\quad\eta_{\sigma\rho}K_{\mu\nu}^{\sp\sp\rho}=0\,,\quad
K_{[\mu\nu]}^{\sp\sp\sp
\rho}=0\,,\quad \ov\nabla_{\mu}\eta_{\nu\rho}=2K_{\mu(\nu\rho)}\,,\quad
\ov\nabla_{\mu}\eta^{\mu\rho}=K^{\rho}\,,\eqno(\new)$$
where round and square brackets denote index symmetrisation and
antisymmetrisation, i.e. $A_{(\mu\nu)}=\textstyle{1\over 2}(A_{\mu\nu}
+A_{\nu\mu})$ and $A_{[\mu\nu]}=\textstyle{1\over 2}(A_{\mu\nu}-A_{\nu\mu})$.
 Using the definition of the first
fundamental tensor (\firstten) and  the properties of the second fundamental
tensor (\tenprop), one can deduce that \eqnam{\conversion}
$$\eqalign{K^{\rho}&=\Vert\gamma\Vert^{-1/2}\partial_a\big{(}\Vert\gamma\Vert^{1/2}
\partial^a X^{\rho}\big{)}+\eta^{\alpha\beta}\Gamma^{\rho}_{\sp
\alpha\beta}\,,\cr K^{\mu\nu\rho}&=\perp_{\sigma}^{\sp
\rho}\big{(}\partial^aX^{\mu} \partial^bX^{\nu}\partial_a\partial_b X^{\sigma}
+ \eta^{\mu\alpha}\eta^{\mu\beta} \Gamma^{\sigma}_{\sp\alpha\beta}
\big{)}\,.\cr}\eqno(\new)$$
using the obvious abbreviation $\partial^a=\gamma^{ab}\partial_b$.
These relations allow one to convert between the  traditional bi--tensor
formalism and the more concise formalism using only background indices.

For an effective action of the simplest type, as exemplified
by the Dirac action for membranes or the Nambu action for strings,
the variational principle gives equations of motion for the brane
that are expressible\refto{car1} simply by \eqnam{\eom}
$$K^{\rho}=0\,,\eqno(\new)$$
in the absence of any coupling of the worldsheet
to any other fields. In this letter, we shall derive the equations of motion
for
small gravitationally induced perturbations in a general gauge. These equations
are shown to simplify in certain special gauge choices. Then we show how the
exact same equations of motion can be deduced by introducing perturbations into
the effective action before using the variational principle. Finally, we use
the
properties of the second fundamental tensor (\tenprop) and the conversion
formulae (\conversion) to compare to those derived using more traditional
notation. This allows direct comparison with results derived in
ref.\refto{Bata}
for gravitational perturbations of a Goto--Nambu string.

\sectbegin{2.}{Gravitationally induced perturbations}

\noindent One of the most fascinating aspects of the Einstein field equations
is
the existence of radiative solutions, similar to those found in
electromagnetism and other gauge theories. This similarity may provide a hint
to
the crucial missing link between general relativity and  gauge theories. In
order
to study such phenomena, one must perturb the Einstein field equations. There
are many different ways of doing this, the most common being  Lagrangian and
Eulerian perturbations. In a Lagrangian scheme, the perturbations are defined
with respect to a reference system that is comoving with the relevant
displacement, whereas in an Eulerian scheme the reference system remains
fixed. Infinitesimal Lagrangian and Eulerian perturbations, denoted by $\dL$
and $\dE$ respectively, can be related by the Lie derivative  ${\cal
L}_{\xi}$,
\eqnam{\lageul}
$$\dL=\dE+{\cal L}_{\xi},,\eqno(\new)$$
with respect to $\xi^{\mu}$ the Lagrangian perturbation of some
arbitrary coordinate system, that is $\xi^{\mu}=\dL x^{\mu}$.
For the purposes of this letter, we shall only consider Lagrangian
perturbations, since $\dL X^{\mu}=0$. However, it is a simple exercise to
deduce the related Eulerian perturbations using (\lageul).

Using the definitions (\firstten) and (\secten), one can deduce that
\eqnam{\varfirstten}
$$\dL\eta^{\mu\nu}= -\eta^{\mu\rho}\eta^{\nu\sigma}\dL
g_{\rho\sigma}\,,\quad\dL
\eta^\mu_{\sp\nu}=\eta^{\mu\rho}\perp^{\sigma}_{\sp\nu} \dL g_{\rho\sigma}\,,
\eqno(\new)$$
and
\eqnam{\varsecten}
$$\dL K_{\mu\nu}^{\sp\sp\rho}=\perp^{\rho}_{\sp\lambda}\eta^{\sigma}_{\sp\mu}
\eta^{\tau}_{\sp\nu}\,\dL\Gamma_{\sp\sigma\tau}^{\lambda}
+\big(2\perp^{\sigma} _{\sp(\mu} K_{\nu)}^{\sp\sp\tau\rho}-K_{\mu\nu}^{\sp\sp
\sigma}\eta^{\tau\rho} \big)\dL g_{\sigma\tau}\,,\eqno(\new)$$
where the Lagrangian variation of the connection is given
by the well known formula
\eqnam{\varconnection}
$$\dL\Gamma_{\sp\sigma\tau}^{\lambda} =g^{\lambda\rho}\big{(}
\nabla_{(\sigma\,}\dL g_{\tau)\rho}-\textstyle{1\over 2}\nabl_{\rho\,}\dL
g_{\sigma\tau} \big{)}\,.\eqno(\new)$$
Previous work on this subject\refto{car3} was restricted to cases for
which the background was fixed in advance, so that there is no Eulerian
variation of the metric, that is $\dE g_{\mu\nu}=0$. In this case,
the Lagrangian variation is just the Lie derivative with respect to
$\xi^{\mu}$,
that is $\dL g_{\sigma\tau}=2\nabl_{(\sigma}\xi_{\tau)}$. Here, we shall allow
also for the possibility that the background spacetime metric is perturbed, so
that there will be a non-zero Eulerian variation, $\dE
g_{\sigma\tau}=h_{\sigma\tau}$. Therefore, the total  Lagrangian variation of
the metric will be given by \eqnam{\totallag}
$$\dL g_{\sigma\tau}=2\nabl_{(\sigma}\xi_{\tau)}+h_{\sigma\tau}\,.\eqno(\new)$$
As with standard treatments of linearized gravity, we shall ignore terms higher
than first order in $h_{\mu\nu}$ and $\xi_{\mu}$.

Using (\varfirstten) and (\varconnection), the Lagrangian variations of the
first fundamental tensor and connection are given by
\eqnam{\finvarfirstten}
$$\dL\eta^{\mu\nu}=-2\eta_\sigma^{\,(\mu}\ov\nabla{^{\nu)}}\xi^\sigma\
-\eta^{\mu\rho}\eta^{\nu\sigma}h_{\rho\sigma}\,,\eqno(\new)$$
and
\eqnam{\finvarconnection}
$$\dL\Gamma_{\sp\sigma\tau}^{\lambda}=\nabl_{(\sigma}\nabl_{\tau)}
\xi^\lambda-\R^\lambda_{\sp(\sigma\tau)\rho}\xi^\rho\ +\nabl_{(\sigma\,}
h_{\tau)}^{\sp\sp\lambda}-{_1\over^2}\nabla^{\lambda}
h_{\sigma\tau}\,,\eqno(\new)$$
where $\R^\lambda_{\sp\sigma\tau\rho}$ is the
background Riemann curvature tensor, which will be negligible in
applications for which the length-scales characterising the geometric features
of interest are small compared with those characterising any background
spacetime curvature. Substituting (\finvarconnection) into (\varsecten) implies
that \eqnam{\finvarsecten}
$$\eqalign{\dL K_{\mu\nu}^{\sp\sp\rho}= &\perp^{\rho}_{\sp\lambda}\big(
\ov\nabl_{(\mu}\ov\nabl_{\nu)}\xi^\lambda-\eta^\sigma_{\sp(\mu}
\eta^\tau_{\sp\nu)}\R^\lambda_{\sp\sigma\tau\alpha}\xi^\alpha-K^\sigma_{\sp(\mu\nu)}
\ov\nabl_\sigma\xi^\lambda\big)\cr &
+\big(2\perp^{\sigma}_{\sp(\mu}K_{\nu)\tau}^{\sp\sp\sp\rho}-\delta^\rho_{\sp\tau}
K_{\mu\nu}^{\sp\sp\sp\sigma}\big)\big(\nabl_\sigma\xi^\tau +
\ov\nabla{^\tau}\xi_\sigma\big)\cr &
\,+\perp^{\rho}_{\sp\lambda}\eta^\sigma_{\sp\mu}\eta^\tau_{\sp\nu}\big(
\nabl_{(\sigma} h_{\tau)}^{\sp\lambda}-\textstyle{1\over 2}\nabla^{\lambda\sp}
 h_{\sigma\tau}\big) +\big(2\perp^{\sigma}_{\sp(\mu} K_{\nu)}^{\sp\sp\tau\rho}
-K_{\mu\nu}^{\sp\sp\sigma}\eta^{\tau\rho}\big)h_{\sigma\tau}\,.}\eqno(\new)$$
The final line is the extra contribution, due to the non-zero Eulerian
perturbation of the metric.
The corresponding expression for Lagrangian perturbations of the curvature
vector is
\eqnam{\varcurv}
$$\dL K^\rho=g^{\mu\nu}\dL K_{\mu\nu}^{\sp\sp\rho} + K_{\mu\nu}^{\sp\sp\rho}
\, \dL g^{\mu\nu}\,.\eqno(\new)$$
Substituting from (\finvarsecten), one finally obtains
\eqnam{\finvarcurv}
$$\eqalign{\dL K^\rho=&\perp^{\rho}_{\sp\lambda}\eta{^{\mu\nu}}
\big(\ov\nabl_\mu\ov\nabl_\nu
\xi^\lambda-\R^\lambda_{\ \mu\nu\sigma}\xi^\sigma\big)
- 2K_{\mu}^{\sp\nu\rho\,} \ov\nabl_\nu \xi^\mu
-K^{\sigma}\big{(}\nabla_{\sigma}\xi^{\rho}+\ov\nabla^{\rho}\xi_{\sigma}\big{)}\cr
&+ \perp^{\rho}_{\sp\lambda}\eta^{\mu\nu}\big( \nabl_{\mu} h_{\nu}^{\sp
\lambda}-\textstyle{1\over 2}\nabla^{\lambda\,}
 h_{\mu\nu}\big) -\big(K^{\mu\nu\rho}+ K^\mu\eta^{\nu\rho}\big)h_{\mu\nu}
\,.}\eqno(\new)$$

All these Lagrangian variations will be invariant
with respect to background coordinate gauge transformations generated by
an arbitrary vector field $\zeta^\rho$, according to the specification
\eqnam{\backgauge}
$$\xi^\rho\mapsto\xi^\rho-\zeta^\rho\,, \quad
h_{\mu\nu}\mapsto h_{\mu\nu}+2\nabl_{(\mu}\zeta_{\nu)}\,.\eqno(\new)$$
The worldsheet itself is also invariant with respect to internal
coordinate gauge transformations generated by an arbitrary tangential
vector field $\epsilon^\rho$ according to the specification
\eqnam{\intgauge}
$$\xi^\rho\mapsto\xi^\rho+\epsilon^\rho\quad
\perp^{\rho}_{\sp\nu}\epsilon^\nu=0\,,\eqno(\new)$$
which can be used to impose the {\it orthogonal gauge}
condition $\eta^\mu_{\ \nu}\xi^\nu=0$,
without restricting the background gauge freedom (\backgauge).
In particular, one may also choose the  the standard {\it harmonic} gauge
condition, $\nabl^{\mu} h_{\mu\nu} -\textstyle{1\over 2}\nabl_{\nu\,}
h=0$,
where $h=g^{\mu\nu}h_{\mu\nu}$,
which is usually the most convenient for practical applications, since
it greatly simplifies the wave equation for $h_{\mu\nu}$. Another possibility
would be to use the comoving gauge
condition $\xi^\rho=0$. However, in this case the wave equation for
$h_{\mu\nu}$ would be much more complicated.
Since $K^{\rho}=0$ for an unperturbed Goto-Nambu string, the equation of motion
for the perturbation $\xi^{\mu}$ is
$$\perp^{\rho\lambda}\eta{^{\mu\nu}}\big{(}\ov\nabl_\mu\ov\nabl_\nu\xi_\lambda-\R
_{\lambda\mu\nu\sigma}\xi^{\sigma}\big{)}-K^{\mu\nu\rho}\big(2\ov\nabl_\nu
\xi_\mu+ h_{\mu\nu}\big) +\perp^{\rho\lambda}\eta^{\mu\nu}\big(\nabl_{\mu}
h_{\nu\lambda} -\textstyle{1\over 2}\nabla_{\lambda} h_{\mu\nu}\big)
=0\,.\eqno(\new)$$

\sectbegin{3.}{The alternative combined perturbation procedure}

\noindent The procedure outlines in the proceeding section consists of making
successive  approximations for the perturbations, whereby one first solves the
zeroth order equation for $X^\mu$ and then solves the first order equation for
$\xi^\mu$. An alternative procedure -- which can be used safely when
$h_{\mu\nu}$
represents a weak, previously given gravitational wave field, but that
leads to runaway solutions for the backreaction problem -- is to choose the
unperturbed world sheet to coincide with the perturbed world sheet.
In this case, the curvature vector non-longer satisfies $K^\rho=0$, but one
now automatically has $\xi^\mu=0$, independently of the
gauge used for $h_{\mu\nu}$. This contrasts with  the successive approximation
approach in which $\xi^\rho$ could only have been set to zero by fixing the
gauge in a manner that would have  been incompatible with the harmonic gauge
condition. Instead of the separate zeroth and first order
equations whereby $K^\rho$  and $\dL K^\rho$ are set to zero
separately, in this alternative procedure one just has a single equation,
expressible -- neglecting second order corrections -- as
$K^\rho+\dL K^\rho=0$, that is
\eqnam{\withk}
$$K^\rho-(K^{\mu\nu\rho}+K^{\mu}\eta^{\nu\rho})h_{\mu\nu}+\perp^{\rho\lambda}
\eta^{\mu\nu}\big(\nabl_{\mu} h_{\nu\lambda} -\textstyle{1\over
2}\nabla_{\lambda} h_{\mu\nu}\big)= 0\,.\eqno(\new)$$

Remembering that $K^{\rho}\sim{\cal O}(h)$, one sees that -- again subject
to neglect of second order corrections, ${\cal O}(h^2)$ -- the tangential
projection of (\withk) is satisfied automatically as a mere identity,
while its projection perpendicular to the worldsheet gives the remaining
non-trivial part of the dynamical equations of the perturbed worldsheet
in the simpler form
\eqnam{\withoutk}
$$K^\rho-K^{\mu\nu\rho}h_{\mu\nu}+\perp^{\rho\lambda}
\eta^{\mu\nu}\big(\nabl_{\mu} h_{\nu\lambda} -\textstyle{1\over
2}\nabla_{\lambda} h_{\mu\nu}\big)= 0\,.\eqno(\new)$$

In this alternative formulation, the inclusion in $h_{\mu\nu}$ of the
gravitational self-field
leads to the familiar problem of unphysical runaway solutions.
This difficulty can be consistently overcome using a local
backreaction approximation and resubstituting the equations of
motion\refto{BSa,Bata}.

\sectbegin{4.}{Variational approach to perturbations}

\noindent One can also obtain the result (\withoutk) by considering the
variation of an action integral of the form
\eqnam{\action}
$$\I= \int \ov\L {\Vert\gamma \Vert}^{1/2} d^{\rm p+1}\sigma=
\int \hat\L {\Vert g \Vert}^{1/2}d^{\rm n}x
 \,,\eqno(\new)$$
where in standard Dirac notation
the distributional background Lagrangian scalar field $\hat \L$ is given
in terms of the regular worldsheet Lagrangian density $\ov \L$ by
\eqnam{\worldlag}
$$ \hat\L=\Vert g \Vert^{-1/2}\int\ov\L \delta^{\rm n}
[x-x\{\sigma\}]{\Vert\gamma \Vert}^{1/2} d^{\rm p+1}\!\sigma
\,.\eqno(\new)$$
In such a formulation, the effect to first order of the
Eulerian variation $g_{\mu\nu}\mapsto  g_{\mu\nu} + h_{\mu\nu}$ will be
expressible simply by $\ov\L\mapsto\ov\Lg$, where the gravitationally coupled
``gross" Lagrangian $\ov\Lg$ is given by
\eqnam{\grosslag}
$$\ov\Lg= \ov\L +\textstyle{1\over 2}h_{\rho\sigma}\ov T{^{\rho\sigma}}\
,\eqno(\new)$$ and the worldsheet energy--momentum density tensor is given
by the standard formula \eqnam{\bart}
$$\ov T{^{\mu\nu}}=2\Vert\gamma\Vert^{-1/2}
{\delta\over\delta g_{\mu\nu}}\Big(\ov\L\Vert\gamma\Vert^{1/2}\Big)
= 2{\delta\ov\L\over\delta g_{\mu\nu}}+\ov\L\eta^{\mu\nu}\,.\eqno(\new)$$
The overhead bar is used to distinguish this regular, fworldsheet
confined tensor field from the corresponding Dirac distributional
background spacetime energy--momentum density tensor field
\eqnam{\hatt}
$$\hat T^{\mu\nu}=2\Vert g\Vert^{-1/2}{\delta\over\delta g_{\mu\nu}}
\Big(\hat\L\Vert g \Vert^{1/2}\Big) =\Vert g\Vert^{-1/2}\int
\ov T{^{\mu\nu}}\, \delta^{\rm n}[x-X\{\sigma^a\}]\, \Vert\gamma \Vert^{1/2}
\, d^{\rm p+1}\sigma\,,\eqno(\new)$$
which will act as the gravitational source for $h_{\mu\nu}$.
Using (\grosslag) one can define the corresponding ``gross" surface energy
momentum density tensor as
\eqnam{\grosstdef}
$$\ov\Tg{^{\mu\nu}}=\ov T{^{\mu\nu}}+\ov\C{^{\mu\nu\rho\sigma}}
h_{\rho\sigma}\ ,\eqno(\new)$$
where $\ov\C{^{\mu\nu\rho\sigma}}=\ov\C{^{\rho\sigma\mu\nu}}$ is the
automatically symmetric {\it hyper-Cauchy tensor}\refto{FS},
\eqnam{\cauchy}
$$ \ov\C{^{\mu\nu\rho\sigma}}=\Vert\gamma\Vert^{-1/2}
{\delta\over\delta g_{\mu\nu}}\Big(\ov T{^{\rho\sigma}}
\Vert\gamma\Vert^{1/2}\Big) ={\delta \ov T{^{\rho\sigma}} \over \delta
g_{\mu\nu}}+{_1\over^2}\ov T{^{\rho\sigma}} \eta^{\mu\nu}\,,\eqno(\new)$$
the relativistic generalisation of the ordinary
space projected Cauchy type elasticity tensor\refto{DeWitt}.

In the application of the variational principle, the worldsheet is
supposed to undergo an infinitesimal virtual displacement $x^\mu\mapsto x^\mu
+\xi^\mu$ so that the action integrand will undergo a
Lagrangian variation  given by
\eqnam{\lagvar}
$$\Vert\gamma\Vert^{-1/2}\dL\big(\ov\Lg\, \Vert\gamma\Vert^{1/2}\big)
=\textstyle{1\over 2}\ov\Tg{^{\mu\nu}}\dL g_{\mu\nu}+\textstyle{1\over 2}\ov
T{^{\mu\nu}} \dL h_{\mu\nu}\ ,\eqno(\new)$$
where the Lagrangian variations of $g_{\mu\nu}$ and $h_{\mu\nu}$ are given by
their Lie derivatives with respect to $\xi^\rho$,
\eqnam{\variations}
$$\dL g_{\mu\nu}=2\nabl_{(\mu}\xi_{\nu)}\,,\quad \dL h_{\mu\nu}=
\xi^\rho\nabl_\rho h_{\mu\nu}+2h_{\rho(\mu}\nabl_{\nu)}\xi^\rho \ .
\eqno(\new)$$
Therefore, the  variation  of the action integrand is
\eqnam{\newlagvar}
$$\Vert\gamma\Vert^{-1/2}\dL\big(\ov\Lg\Vert\gamma\Vert^{1/2}\big)
= \ov\nabl_\mu\Big(\xi^\rho(\ov \Tg{^{\mu}}_{\rho}+\ov T{^{\mu\nu}}
h_{\nu\rho})\Big)-\xi^\rho\Big(\ov\nabl_\mu\big(\ov\Tg{^{\mu}}_{\rho}
+\ov T{^{\mu\nu}}h_{\nu\rho}\big)-\textstyle{1\over 2}\ov
T{^{\mu\nu}}\nabl_\rho h_{\mu\nu}\Big) \,.\eqno(\new)$$
The first term is a surface divergence and can be ignored by using Green's
theorem. Therefore, variational invariance reduces to
the requirement  that the coefficient of $\xi^\rho$ in the second term
should vanish, that is
\eqnam{\vareom}
$$\ov\nabl_\mu\big(\ov\Tg{^{\mu\rho}}+\ov T{^{\mu\nu}}h_{\nu}^{\sp\rho}
\big)=\textstyle{1\over 2}\ov T{^{\mu\nu}}\nabl^\rho
h_{\mu\nu}\,.\eqno(\new)$$ Regrouping the first order terms onto the right
side, this dynamical equation can conveniently be rewritten as
\eqnam{\acteom}
$$\ov\nabl_\mu\ov T{^{\mu\rho}}=\ov f{^\rho}\ ,\eqno(\new)$$
where the effective surface force density due to the gravitational
perturbation is given by
\eqnam{\forcedensity}
$$\ov f{^\rho}=\textstyle{1\over 2}\ov T{^{\mu\nu}}\nabla^\rho h_{\mu\nu}-
\ov\nabl_\mu\big(\ov T{^{\mu\nu}}h_\nu{^\rho}+\ov\C{^{\mu\rho\nu\lambda}}
h_{\nu\lambda}\big)\,.\eqno(\new)$$

For a Dirac--Goto--Nambu
membrane or string, the unperturbed Lagrangian is just a constant and therefore
\eqnam{\diracnamgoto}
$$\ov\L=-m^{\rm p+1}\,,\quad \ov T{^{\mu\nu}}=-m^{\rm p+1}\eta^{\mu\nu}
\,,\eqno(\new)$$
where $m$ is some fixed parameter having the dimensions of a mass.
In general $m$ would be expected to be of the same order of
magnitude as the relevant Higgs mass of the underlying field theoretical
model. However, in some cases it could be
very much larger, for example a global string. Using (\diracnamgoto) one can
deduce that the hyper-Cauchy tensor is
\eqnam{\actcauchy}
$$\ov\C{^{\mu\nu\rho\sigma}}=m^{\rm
p+1}\big(\eta^{\mu(\rho}\eta^{\sigma)\nu}-\textstyle{1\over 2}
\eta^{\mu\nu}\eta^{\rho\sigma}\big)\,.\eqno(\new)$$
If one now substitutes (\diracnamgoto) and (\actcauchy) into (\grosslag) and
(\grosstdef), the gravitationally coupled ``gross" Lagrangian is given by
\eqnam{\actlag}
$$\ov\Lg=-m^{\rm p+1}\big(1+\textstyle{1\over
2}\eta^{\mu\nu}h_{\mu\nu}\big) \,.\eqno(\new)$$
and the corresponding ``gross"
surface energy--momentum tensor is given by
\eqnam{\actt}
$$\ov \Tg{^{\mu\nu}}=m^{\rm
p+1}\bigg{[}\eta^{\mu\rho}\eta^{\nu\sigma}h_{\rho\sigma}-\eta^{\mu\nu}
\big(1+\textstyle{1\over 2}\eta^{\rho\sigma}h_{\rho\sigma}\big)\bigg{]}
\,.\eqno(\new)$$ Using the properties the first and second fundamental tensors
(\tenprop), one can deduce that the force density induced by the gravitational
perturbations is $$\ov f^\rho=m^{\rm
p+1}\bigg{[}\perp^{\rho\lambda}\eta^{\mu\nu}\big{(}\nabla_{\mu}h_{\nu\lambda}-
\textstyle{1\over
2}\nabla_{\lambda}h_{\mu\nu}\big{)}
+\big{(}\perp^{\rho\nu}K^{\mu}+\textstyle{1\over
2}\eta^{\mu\nu}K^{\rho}-K^{\mu\nu\rho}\big{)}h_{\mu\nu}\bigg{]}\,.\eqno(\new)$$
This force is orthogonal to the worldsheet, that is $\eta^{\mu}_{\sp\rho}\ov
f^{\rho}=0$, which is a Noether identity resulting from the lack of internal
structure in the Dirac--Goto--Nambu case.

Using (\acteom), the dynamical equations are thus obtained in the final form
\eqnam{\dyneom}
$$K^\rho+\big{(}\perp^{\rho\nu}K^{\mu}+\textstyle{1\over
2}\eta^{\mu\nu}K^{\rho}-K^{\mu\nu\rho}\big{)}h_{\mu\nu}+\perp^{\rho\lambda}
\eta^{\mu\nu}\big{(}\nabla_{\mu}h_{\mu\lambda}-\textstyle{1\over
2}\nabla_{\lambda}h_{\mu\nu}\big{)}=0\,.\eqno(\new)$$
In order to account for the small,second order discrepancy between this
final variational equation (\dyneom) and the previous equation (\withoutk)
that was obtained via a less sophisticated approach by considering
$K^{\rho}+\dL K^{\rho}=0$, one must understand that in the variational
case we are effectively considering variations of
$\Vert\gamma\Vert^{1/2} K_{\rho}$, instead of just $K^{\rho}$.
The equation (\dyneom) can be seen to be exactly equivalent to
\eqnam{\pertequation}
$$K^{\rho}+\dL K^{\rho}+\Vert\gamma\Vert^{-1/2}\dL\big{(}
\Vert\gamma\Vert^{1/2}\big{)}K^{\rho}+h^{\rho}_{\sp\sigma}K^{\sigma}
=0\,\eqno(\new)$$
where
$\Vert\gamma\Vert^{-1/2}\dL\big{(}\Vert\gamma\Vert^{1/2}\big{)}=\textstyle
{1\over 2}\eta^{\mu\nu}h_{\mu\nu}$.
Another alternative equation could be similarly deduced by considering
variations of
$\Vert\gamma\Vert^{1/2}K^{\rho}$. However all these different dynamical
equations can be seen to agree to within corrections ${\cal O}(h^2)$,
when one uses the fact that $K^{\rho}\sim{\cal O}(h)$.

\sectbegin{5.}{Comparison to results in traditional notation}

\noindent We have shown that the equation (\withoutk)
describes the dynamics of a perturbed Dirac Goto-Nambu membrane or
string. This problem has also been studied for strings  using more
traditional notation\refto{Bata}. It should be possible to show that the
results obtained by the two approaches are ultimately equivalent. In order
 to do this we split
the forcing terms up into two parts, writing the dynamical equation
(\withoutk) in the form
\eqnam{\decomp}
$$K^{\rho}=F_1^{\rho}+F_2^{\rho}\, ,\eqno(\new)$$
with
\eqnam{\decoterms}
$$F_1^{\rho}=-\perp^{\rho\lambda}\eta^{\alpha\beta}\bigg{(}\nabla_{\alpha}
h_{\lambda\beta}-\textstyle{1\over 2}\nabla_{\lambda}h_{\alpha\beta}\bigg{)}
\, , \hskip 1 cm F_2^{\rho}=h_{\mu\nu}K^{\mu\nu\rho}\, ,\eqno(\new)$$
in which  $F_1^{\rho}$  depends only  on derivatives of $h_{\mu\nu}$ and
$F_2^{\rho}$ depends only on the undifferentiated metric
perturbation $h_{\mu\nu}$.

Using the definition of $\perp^{\mu\nu}$, it is a trivial exercise to show
that
\eqnam{\actfone}
$$\eqalign {F_1^{\rho}=&
\eta^{\rho\lambda}\eta^{\alpha\beta}\bigg{(}\nabla_{\alpha}h_{\lambda\beta}
-\textstyle{1\over
2}\nabla_{\lambda}h_{\alpha\beta}\bigg{)}-\eta^{\alpha\beta}\bigg{(}\nabla_{\alpha}h^{\rho}_{\sp\beta}
-\textstyle{1\over
2}\nabla^{\rho}h_{\alpha\beta}\bigg{)}\cr=&
\bigg{(}\eta^{\alpha\gamma}\eta^{\beta\rho}-\textstyle{1\over
2}\eta^{\alpha\beta}\eta^{\gamma\rho}\bigg{)}\nabla_{\gamma}h_{\alpha\beta}
-\textstyle{1\over2}\eta^{\alpha\beta}\bigg{(}-\nabla^{\rho}h_{\alpha\beta}
+\nabla_{\alpha}h^{\rho}_{\sp\beta}+\nabla_{\beta}h_{\alpha}^{\sp\rho}\bigg{)}
\, .} \eqno(\new)$$
Evaluating the above expression for $F_2^{\rho}$ is more tricky.
Using the formula for $\ov\nabla_{\alpha}\eta^{\mu\nu}=
2K_{\alpha}^{\sp(\mu\nu)}$ with
the formula (\firstten) for the first fundamental tensor,
and making the chain rule substitution $\partial_a\eta^{\mu\nu}=
\partial_aX^{\alpha}\partial_{\alpha}\eta^{\mu\nu}$, one can deduce that
\eqnam{\newfone}
$$\eqalign{F_2^{\rho}=&
h_{\mu\nu}\bigg{(}
\eta^{\mu\alpha}\partial_{\alpha}\eta^{\rho\nu}-\textstyle{1\over
2}\eta^{\alpha\rho}\partial_\alpha\eta^{\mu\nu}+
\Gamma^{\rho}_{\sp\alpha\beta}\eta^{\alpha\mu}\eta^{\beta\nu}\bigg{)}
\cr=&
h_{\mu\nu}\bigg{(}\partial_{\alpha}\eta^{\rho\nu}\gamma^{ab}\partial_aX^{\alpha}
\partial_bX^{\nu}-\textstyle{1\over
2}\partial_\alpha\eta^{\mu\nu}\gamma^{ab}\partial_aX^{\alpha}\partial_bX^{\rho}
+\Gamma^{\rho}_{\sp\alpha\beta}
\eta^{\alpha\mu}\eta^{\beta\nu}\bigg{)}
\, ,}\eqno(\new)$$
from which we finally obtain
\eqnam{\oldterm}
$$
F_2^{\rho}=h_{\mu\nu}\bigg{(}\gamma^{ab}\partial_a\big{(}\gamma^{cd}\partial_cX^{\rho}
\partial_dX^{\mu}\big{)}\partial_bX^{\nu}-\textstyle{1\over
2}\gamma^{ab}\partial_a\big{(}\gamma^{cd}\partial_cX^{\mu}\partial_dX^{\nu}
\big{)}\partial_bX^{\rho}+\Gamma^{\rho}_{\sp\alpha\beta}
\eta^{\alpha\mu}\eta^{\beta\nu}\bigg{)}\,.\eqno(\new)$$
It can be checked that this agrees with what is obtained
by the traditional approach\refto{Bata}.

\nosectbegin{Acknowledgments}

\noindent We would like to thank Patricio Letelier, Paul Shellard
and Xavier Martin for helpful discussions. RB acknowledges the support of the
Engineering and Physical Sciences Research Council. We also acknowledge the
British Council Alliance Exchange grant for funding the visit to Meudon, where
this work was initiated.



\def\hang{}


\def\jnl#1#2#3#4#5#6{\hang{#1 [#2], {\it #4\/} {\bf #5}, #6.}
									}


\def\jnlerr#1#2#3#4#5#6#7#8{\hang{#1 [#2], {\it #4\/} {\bf #5}, #6.
{Erratum:} {\it #4\/} {\bf #7}, #8.}
									}


\def\prep#1#2#3#4{\hang{#1 [#2],`#3', #4.}
									}

\def\proc#1#2#3#4#5#6{\hang{#1 [#2], in {\it #4\/}, #5, eds.\ (#6).}
}
\def\procu#1#2#3#4#5#6{\hang{#1 [#2], in {\it #4\/}, #5, ed.\ (#6).}
}

\def\book#1#2#3#4{\hang{#1 [#2], {\it #3\/} (#4).}
									}



\def\prl{Phys.\ Rev.\ Lett.}
\def\pr{Phys.\ Rev.}

\def\prp{Phys.\ Rep.}

\def\apj{Ap.\ J.}

\def\cqg{Class.\ Quant.\ Grav.}

\def\mn{M.$\,$N.$\,$R.$\,$A.$\,$S.}

\def\cup{Cambridge University Press}

\def\skip{\vskip -4pt}

\nosectbegin{References}

\references

\baselineskip 10pt
\let\it=\nineit
\let\rm=\ninerm
\let\bf=\ninebf
\rm

\refis{BSa}
\prep{Battye, R.A., \& Shellard, E.P.S.}{1994}{String radiative
backreaction }{DAMTP Preprint R94/31, to appear {\it Phys. Rev.
Lett}}\prep{Battye, R.A., \& Shellard, E.P.S.}{1994}{Radiative backreaction on
global strings}{DAMTP Preprint R94/50, to appear {\it\pr} {\bf D}}\skip

\refis{Bata}
\prep{Battye, R.A.}{1995}{Gravitational radiation backreaction on
cosmic strings}{DAMTP Preprint R95/18, Submitted to {\it Class. Quant.
Grav}}\skip

\refis{ASa}
\jnl{Allen, B., \& Shellard, E.P.S.}{1990}{Cosmic string evolution---a
numerical simulation}{\prl}{64}{119} \proc{Shellard, E.P.S., \& Allen,
B.}{1990}{On the evolution of
cosmic strings}{Formation and Evolution of Cosmic Strings}{Gibbons, G.W.,
Hawking, S.W., \& Vachaspati, T.}{\cup} \skip

\refis{FS}
\jnl{Friedman, J., \& Schutz B.F.,}{1975}{}{\apj}{200}{205}\skip

\refis{DeWitt}
\procu{De Witt, B.}{1962}{}{Gravitation}{L. Witten}{Wiley, New York}\skip

\refis{ASb}
\jnl{Allen, B., \& Shellard, E.P.S.}{1992}{Gravitational radiation from a
cosmic string network}{\pr}{D45}{1898}\skip

\refis{structure}
\jnl{Zel'dovich, Ya.B.}{1980}{Cosmological fluctuations produced
near a singularity}{\mn}{192}{663}
\jnlerr{Vilenkin, A.}{1981}{Cosmological density fluctuations produced by
vacuum
strings}{\prl}{46}{1169}{46}{1496}\skip

\refis{BB}
\jnl{Bennett, D.P., \& Bouchet, F.R.}{1991}
{Constraints on the gravity wave background generated by cosmic
strings}{\pr}{D43}{2733}\skip

\refis{AC}
\jnl{Caldwell, R.R., \& Allen, B.}{1992}{Cosmological
constraints on cosmic string gravitational radiation}{\pr}{D45}{3447}\skip

\refis{kibble}
\jnl{Kibble, T.W.B.}{1980}{Some implications of a cosmological phase
transition}{\prp}{67}{183}\skip

\refis{car1}
\jnl{Carter, B.}{1989}{}{Phys. Lett.}{B228}{466}
\proc{Carter, B.}{1990}{On the evolution of cosmic strings}{Formation and
Evolution of Cosmic Strings}{Gibbons, G.W., Hawking, S.W., \& Vachaspati, T.}
{\cup}\skip

\refis{car2}
\jnl{Carter, B.}{1992}{}{J. Geom. and Phys.}{8}{53}
\jnl{Carter, B.}{1992}{}{\cqg}{9S}{19}\skip

\refis{car3}
\jnl{Carter, B.}{1993}{}{\pr}{D48}{4835}\skip

\refis{CMS}\jnl{Carter, B., Martin, X, \& Sakellariadou, M}{1994}{}
{\pr}{D50}{682}\skip

\refis{NATOVS}
\book{Vilenkin, A., \& Shellard, E.P.S.}{1994}{Cosmic strings and other
topological defects}{\cup} \book{Davis, A.C., \& Brandenberger,
R.H.}{1995}{Formation and interactions of topological defects}{Plenum
Press}\skip

\endreferences

\vfill
\end